\documentclass[review,12pt]{elsarticle}

\usepackage{amssymb}

\journal{Physics Letters A}
\usepackage{amsmath}
\usepackage{amsfonts}
\usepackage{geometry}                
\usepackage[parfill]{parskip}    
\usepackage{graphicx}
\usepackage{amssymb}
\usepackage{epstopdf}
\usepackage{psfrag}

\newcommand {\be}{\begin{equation}}
\newcommand {\bes}{\begin{displaymath}}
\newcommand {\es}{\end{displaymath}}
\newcommand {\e}{\end{equation}}

\newcommand {\bea}{\begin{eqnarray}}
\newcommand {\ea}{\end{eqnarray}}

\usepackage{xspace}

\usepackage{subfigure}

\newcommand{\secref}[1]{{Section~\ref{#1}}}

\begin{document}

\begin{frontmatter}


\title{Localized modulated wave solutions in diffusive glucose-insulin systems}

\author[Byd1,cetic]{Alain Mvogo\corref{cor1}}
\ead{mvogal$_{-}$2009@yahoo.fr} \cortext[cor1]{Corresponding author}
\address[Byd1]{Laboratory of Biophysics, Department of Physics, Faculty of Science, University of Yaounde I, P.O. Box 812, University of Yaounde, Cameroon}
\address[cetic]{Centre d'Excellence Africain en Technologies de l'Information et de la Communication, University of Yaounde I, Cameroon}

\author[at,atb]{Antoine Tambue}
\address[at]{The African Institute for Mathematical Sciences (AIMS) and Stellenbosch University, 6-8 Melrose Road, Muizenberg 7945, South Africa}
\address[atb]{Center for Research in Computational and Applied Mechanics (CERECAM), and Department of Mathematics and Applied Mathematics, University of Cape Town, 7701 Rondebosch, South Africa.}

\author[cetic,Nyd1]{Germain H. Ben-Bolie}
\address[Nyd1]{Laboratory of Nuclear Physics, Department of Physics, Faculty of Science, University of Yaounde I, P.O. Box 812, University of Yaounde, Cameroon}

\author[cetic,Myd1]{Timol\'{e}on C. Kofan\'{e}}
\address[Myd1]{Laboratory of Mechanics, Department of Physics, Faculty of Science, University of Yaounde I, P.O. Box 812, University of Yaounde, Cameroon}
\address{}

\begin{abstract}
\qquad  We investigate intercellular insulin dynamics in an array of
diffusively coupled pancreatic islet $\beta$-cells. The cells are
connected via gap junction coupling, where nearest neighbor
interactions are included. Through the multiple scale expansion in
the semi-discrete approximation, we show that the insulin dynamics
can be governed by the complex Ginzburg-Landau equation. The
localized solutions of this equation are reported. The results
suggest from the biophysical point of view that the insulin
propagates in pancreatic islet $\beta$-cells using both temporal and
spatial dimensions in the form of localized modulated waves.

\end{abstract}
\end{frontmatter}

\section{Introduction}
\label{sec:intr}

Blood glucose levels are controlled by a complex interaction of
multiple chemicals and hormones in the body. The metabolism of
glucose into the $\beta$-cells leads to increase in the adenosine
triphosphate concentration, closure of ATP-sensitive $K^{+}$
channels, depolarization of the $\beta$-cell membrane and opening of
the voltage-dependant $C_{a}^{2+}$ channels, thereby allowing
$C_{a}^{2+}$ influx \cite{Ashcroft}. The resultant rise in
intracellular $C_{a}^{2+}$ concentration in the $\beta$-cell
triggers insulin secretion. Insulin, which is secreted from
pancreatic $\beta$-cells is the key hormone regulating glucose
levels.

Physiological responses generated within a cell can propagate to
neighboring cells through intercellular communication involving the
passage of a molecular signal to a bordering cell through a gap
junction \cite{Isakson,Osipchuk}, through extracellular
communication involving the secretion of molecular signals
\cite{Isakson} such as hormones, neurotransmitters, etc., and also
trough extracellular calcium signaling \cite{Hofer1,Gracheva}.
Oscillations of $C_{a}^{2+}$ rather than metabolism in the
$\beta$-cell are thought to be the direct cause of these
oscillations in insulin secretion \cite{Gilon}. Sneyd \emph{et al}.
\cite{Sneyd} proposed a dynamical model of such oscillations, which
assume gap junction diffusion of inositol-1,4,5-triphosphate between
adjacent cells. The diffusion of inositol-1,4,5-triphosphate between
cells then initiate not only $C_{a}^{2+}$ oscillations but also
insulin oscillations in adjacent cells.

It is well known that the dynamics of the insulin is very relevant
because it is related to the onset of the pathologies  such as
diabetes caused by elevated blood glucose. Careful diabetes mellitus
self-management is essential in avoiding chronic complications that
compromise health, and is characterized by many and often not
readily observable clinical effects \cite{Diabete1}. Along the same
line, attention has been paid with the increased emphasis on
derangements of the sensitivity of tissues to insulin in diverse
pathological conditions like diabetes, obesity and cardiovascular
diseases \cite{Gaetan,Defronzo,Frontoni}. Therefore, there is an
urgent need for improved diagnostic methods that provide more
precise clinical assessments and sensitive detection of symptoms at
earlier stage of the disease. This may be facilitated by improved
mathematical models and tools related to interrelationship dynamics
among physiological variables implicated in the glucose-insulin
system. This assumption motivates the present work, where a
diffusive model of coupled pancreatic islet $\beta$-cells is
investigated. The cells are connected via the gap junction coupling
which consists of a mechanism used by cells to coordinate and
synchronize their information \cite{Sneyd}. We then investigate a
clear analytical solution describing the dynamics of insulin in the
system. By means of the multiple scale expansion in the
semi-discrete approximation, we use the Li\'{e}nard form of a
diffusive system to obtain the complex Ginzburg-Landau (CGL)
equation, which describes the evolution of modulated waves in this
system. We obtain an expression of the hormonal wave by using of the
envelope soliton solution of the CGL by Pereira  and Stenflo
\cite{Stenflo}, and Nozaki and Bekki \cite{Nozaki}. The solution
reveal that the hormonal wave is a localized nonlinear solution
which propagates in the form of a breather-like coherent structure.

The rest of the paper is organized as follows: in \secref{sec2}, we
present the mathematical model. In \secref{sec3}, we find an
envelope soliton in the model by applying the multiple scale
expansion in the semi-discrete approximation. Our work is summarized
in \secref{sec4}.

\section{Mathematical model}
 \label{sec2}
We propose in this paper a diffusive model of coupled pancreatic
islet $\beta$-cells, in which the cells are connected via the gap
junction coupling. Diffusive cell models with gap junction coupling
are quite interesting in describing and characterizing quasi-perfect
intercellular communication \cite{Sneyd}. For the mathematical
modeling, let us consider $x_{n}$ and $y_{n}$ as the intracellular
concentrations of glucose and insulin in the $n$th cell,
respectively. The equations of $N$ coupled cells in the system
through the gap junction are given by
\begin{equation}\label{eq1}
\dot{x}_{n}=-a_{1}x_{n}-a_{2}x_{n}y_{n}+a_{3},
\end{equation}
\begin{equation}\label{eq2}
\dot{y}_{n}=D(y_{n+1}-2y_{n}+y_{n-1})+b_{1}x_{n}-b_{2}y_{n},
\end{equation}
where  $n = 1,...,N$. The parameter $a_{1}$ is the rate constant
which represents insulin-independent glucose disappearance, $a_{2}$
is the rate constant which represents insulin-dependent glucose
disappearance. In other terms, $a_{2}$ describes the modulation of
the effective kinetic constant of the glucose utilization by insulin
action. The parameter $a_{3}$ is the glucose infusion rate, $b_{1}$
is the rate constant which represents insulin production due to
glucose stimulation and $b_{2}$ is the rate constant which
represents insulin degradation. The parameter $D$ is the coupling
strength of the gap junction. We have considered two nearest
neighbors coupling in a weak coupling regime. A weak coupling
between neighboring cells is a situation that arises in the study of
bursting activity in the $\beta$-cell islets of the pancreas, which
secrete insulin in response to glucose level
\cite{Raghavachari,Perez}. The model also predicts that oscillations
occur if there is sufficient diffusion ($D>0.1$) to create adequate
concentrations mixing in the reacting layers of the cells
\cite{Keener}.

Note that the model is nonlinear, due to the presence of the
bilinear term between the internal variable $y(t)$ representing the
insulin action and the variable $x(t)$ representing in the first
equation the glucose concentration. The parameter values chosen are
those from the literature of healthy volunteer undergoing the intra
venous glucose tolerance test. In a clinical experiment conducted
and reported in Ref. \cite{Gaetano}, ten healthy volunteers (5 males
and 5 females) participated in the study. As indicated  by Gaetano
and Arino \cite{Gaetano}, all subjects had negative family and
personal histories for diabetes mellitus and other endocrine
diseases, were on no medications, had no current illness and had
maintained a constant body weight for the six months preceding the
study.   In the present paper, we have taken the data of the two
first subjects as listed in the Table 1.

\begin{table}[h!]
\begin{center}
\begin{tabular}{|c|c|c|c|c|c|c|}
\hline
$\text{Subjects}$ & $a_{1} $& $ a_2 $ &  $a_3 $ & $b_1 $& $ b_{2}$ \\
\hline
1 & $0.0226$ & $3.8\times10^{-8}$  & $1.56$   & 0.0022  &  $0.0437$ \\
\hline
2 & $0.0509$ & $1.29\times10^{-7}$  & $4.02$   & 0.0096  &  $0.2062$ \\
\hline
\end{tabular}
\end{center}
\caption{Parameter values \cite{Gaetano}.} \label{phi}
\end{table}

Many papers have been published detailing the occurrence mode
oscillations of pancreatic islet $\beta$-cells. In the present
paper, the system being nonlinear, we are mainly interested by
nonlinear solutions that can describe the nonlinear dynamics of
insulin. It is convenient to transform the system into the wave
form. This is achieved by differentiating the second equation and
substituting $x_{n}$ into the obtained second-order ordinary
differential equation. The above transformations did not
fundamentally affect the structure of the system, but allow us to
conveniently write Eqs. \eqref{eq1} and \eqref{eq2} in a Li\'{e}nard
form that is a second-order differential equation with a small
damping term. The governing equation for the insulin concentration
then reads
\begin{equation}\label{eq3}
\begin{split}
\ddot{y}_{n}+\Omega_{0}^{2}y_{n}&+(\nu_{0}+\nu_{1}y_{n})\dot{y}_{n}
+\lambda_{1}y_{n}^{2}+K
=D_{0}(y_{n+1}-2y_{n}+y_{n-1})\\
&+D_{1}(\dot{y}_{n+1}-2\dot{y}_{n}+\dot{y}_{n-1})+D_{2}(y_{n+1}-2y_{n}+y_{n-1})\dot{y}_{n},
\end{split}
\end{equation}
where $\Omega_{0}^{2}=a_{1}b_{2}$, $\nu_{0}=a_{1}+b_{2}$,
$\nu_{1}=a_{2}$, $\lambda_{1}=a_{2}b_{2}$, $D_{0}=a_{1}D$,
$D_{1}=D$, $D_{2}=a_{2}D$ and $K=-b_{1}a_{3}$.

For such equations, perturbation approaches are used to obtain
nearly exact solutions. Accordingly, we introduce a new variable
$\psi_{n}$ such that
\begin{equation}\label{eq4}
y_{n}=\epsilon \psi_{n}.
\end{equation}
As we are interested  for solution in a weakly dissipative medium,
we assume the parameters $\nu_{0}$ and $D_{1}$ perturbed at the
order $\epsilon^{2}$. Keeping the first nonlinear term of the
development,  Eq. \eqref{eq3} reads
\begin{equation}\label{eq5}
\begin{split}
\ddot{\psi}_{n}+\Omega_{0}^{2}\psi_{n}&+(\epsilon^{2}\nu_{0}+\epsilon\nu_{1}\psi_{n})\dot{\psi}_{n}
+\epsilon\lambda_{1}\psi_{n}^{2}
=D_{0}(\psi_{n+1}-2\psi_{n}+\psi_{n-1})\\
&+\epsilon^{2}D_{1}(\dot{\psi}_{n+1}-2\dot{\psi}_{n}+\dot{\psi}_{n-1})+\epsilon
D_{2}(\psi_{n+1}-2\psi_{n}+\psi_{n-1})\dot{\psi}_{n}.
\end{split}
\end{equation}

The solutions of this equation are regarded up of carrier waves
modulated by envelope signal, called envelope solitons which appears
naturally for most weakly dispersive and nonlinear systems in the
small amplitude limit \cite{Remoissenet}. In the next section, the
multiple scale expansion in the semi-discrete approximation is used
to find the envelope soliton solution of Eq. \eqref{eq3}. The method
has been found a powerful tool  in solving similar equation
\cite{Remoissenet}.

\section{Multiple scale expansion in the
semi-discrete approximation}
 \label{sec3}
Referring to the multiple scale expansion \cite{Kaup1,Kaup2}, we
proceed by making a change of variables according to the  space and
new time scales $Z_n=\epsilon^{n}z$ and $T_n=\epsilon^{n}t$,
respectively. That is we are looking for solution $y(z,t)$ depending
on these new set of variables as a perturbation series of functions
\begin{equation}\label{eq6}
y(z,t) = \sum\limits_{n = 1}^\infty  {{\epsilon^n}{\psi_n}\left(
{{Z_0},{Z_1},{Z_2},...,{T_0},{T_1},{T_2},...} \right)},
\end{equation}
where $Z_n$ and $T_n$ are treated as independent variables.
\subsection{Equation of motion of the amplitude}
 \label{sec3.1}
The semi-discrete approximation is a perturbation technique in which
the carrier waves are kept discrete while the amplitude is treated
in the continuum limit. For this, we look for modulated wave
solution of the form
\begin{equation}\label{eq7}
\begin{split}
\psi_{n}=&F_{1n}e^{i\theta_{n}}+F_{1n}^{*}e^{-i\theta_{n}}
+\epsilon[F_{0n}+(F_{2n}e^{2i\theta_{n}}+F_{2n}^{*}e^{-2i\theta_{n}})]+O(\epsilon^{2}),
\end{split}
\end{equation}
where $\theta_{n}=qn-\omega t$, $q$ is the wave vector and $\omega$
is the frequency. The procedure will consist of replacing the form
of the solution Eq. \eqref{eq7} and its derivatives in the different
terms of Eq. \eqref{eq5}. We then group terms in the same power of
$\epsilon$, which leads us to a system of equations. Each of those
equations will correspond to each approximation for specific
harmonics.

Substituting Eq. \eqref{eq7} in Eq. \eqref{eq5} gives
\begin{equation}\label{eq8}
\begin{split}
&(\ddot {F}_{1,n}-2i\omega\dot{F}_{1,n}-\omega^{2}
F_{1,n})e^{i\theta_{n}}+\epsilon (\ddot{F}_{0,n}) +\epsilon
(\ddot{F}_{2,n}-4i\omega \dot{F}_{2,n}-4\omega^{2}F_{2,n}
)e^{2i\theta_{n}}\\
&+\Omega_{0}^{2}[F_{1,n}e^{i\theta_{n}}+\epsilon F_{0,n}+\epsilon
F_{2,n}e^{2i\theta_{n}}] +\epsilon^{2} \nu_{0}
(\dot{F}_{1,n}-i\omega F_{1,n})e^{i\theta_{n}}+\epsilon
\nu_{1}[F_{1,n}(\dot{F}_{1,n}-i\omega
F_{1,n})e^{2i\theta_{n}}\\
&+\epsilon F_{1,n} \dot{F}_{0,n}e^{i\theta_{n}}+\epsilon
F_{0,n}(\dot{F}_{1,n}-i\omega F_{1,n})e^{i\theta_{n}}] +\epsilon
\lambda_{1} [(F_{1,n}^{2}e^{2i\theta_{n}}+2 F_{1,n} F_{1,n}^{*})\\
&=+2\epsilon^{2} \lambda_{1}(F_{1,n} F_{0,n}+F_{1,n}^{*}
F_{2,n})e^{i\theta_{n}}]+D_{0}(F_{1,n+1}e^{iqa}+F_{1,n-1}e^{-iqa}-2F_{1,n})e^{i\theta_{n}}\\
&+\epsilon D_{0}(F_{0,n+1}+F_{0,n-1}-2F_{0,n})+\epsilon D_{0}(F_{2,n+1}e^{2iqa}+F_{2,n-1}e^{-2iqa}-2F_{2,n})e^{2i\theta_{n}}\\
&+\epsilon^{2}D_{1}[\dot{F}_{1,n+1}
e^{iqa}+\dot{F}_{1,n-1}e^{-iqa}-2\dot{F}_{1,n}-i\omega(F_{1,n+1}e^{iqa}+F_{1,n-1}e^{-iqa}-2
F_{1,n})]e^{i\theta_{n}} \\
&+\epsilon D_{2}F_{1,n}
(F_{1,n+1}e^{iqa}+F_{1,n-1}e^{-iqa}-2F_{1,n})e^{2i\theta_{n}}+\epsilon^{2}
D_{2}F_{1,n}(F_{0,n+1}+F_{0,n-l}-2F_{0,n})e^{i\theta_{n}}\\
&+\epsilon^{2} D_{2}F_{0,n}
(F_{1,n+1}e^{iqa}+F_{1,n-1}e^{-iqa}-2F_{1,n})e^{i\theta_{n}}.
\end{split}
\end{equation}

Since the envelope function varies slowly in space and time, we use
the continuum approximation for $F$ in a multiple scale expansion
such that
\begin{equation}\label{eq9}
F_{n\pm1} =F \pm \epsilon\frac{\partial F}{{\partial {Z_1}}} \pm
\epsilon^{2} \frac{\partial F}{{\partial {Z_2}}} + {
\frac{\epsilon^2}{2}}\frac{\partial^{2} F}{{\partial {Z_1^{2}}}} +
O(\epsilon^{3}),
\end{equation}
and
\begin{equation}\label{eq10}
\frac{\partial F_{n}}{\partial t}= \epsilon \frac{\partial
F}{\partial {T_1}} + \epsilon^2\frac{\partial^{2} F}{\partial {T_2}}
+ O(\epsilon^3).
\end{equation}

Equating dc, first-, and second-harmonic terms, we get respectively
\begin{equation}\label{eq12}
F_{0}=-\frac{2\lambda_{1}}{\Omega_{0}^{2}}|F_{1}|^{2},
\end{equation}
\begin{equation}\label{eq13}
F_{2}=\frac{\lambda_{1}-i\omega\nu_{1}-4D_{2}\sin^{2}(\frac{q}{2})}{3\Omega_{0}^{2}+16D_{0}\sin^{4}(\frac{q}{2})}F_{1}^{2},
\end{equation}
\begin{equation}\label{eq11}
\begin{split}
\frac{\partial^2 F_{1}}{\partial T_1^2}&-2i\omega \frac{\partial
F_{1}}{\partial T_2}= i\omega \nu_{0}F_{1}
+(i\omega\nu_{1}-2\lambda_{1})\Big[\frac{-2\lambda_{1}}{\Omega_{0}^{2}}
+\frac{\lambda_{1}-i\omega\nu_{1}-4D_{2}\sin^{2}(\frac{q}{2})}{3\Omega_{0}^{2}+16D_{0}\sin^{4}(\frac{q}{2})}\Big]|F_{1}|^2F_{1}\\
&+2iD_{0}\sin(q)\frac{\partial F_{1}}{\partial
Z_2}+D_{0}\cos(q)\frac{\partial^2 F_{1}}{\partial Z_1^2}+4i\omega
D_{1}\sin^{2}(\frac{q}{2})F_{1}+\frac{8\lambda_{1}D_{2}}{\Omega_{0}^{2}}
|F_{1}|^2F_{1}.
\end{split}
\end{equation}

In the above calculation, we have used the dispersion relation for
the carrier wave
\begin{equation}\label{eq14}
\omega^{2}=\Omega_{0}^{2}+4D_{0}\sin^{2}(\frac{q}{2}),
\end{equation}
obtained by linearizing Eq.(8). As we observe in Fig. (1), the
corresponding linear spectrum for the first two subjects of Ref.
\cite{Gaetano} is related to the system parameters. However for the
parameter values related to the subject 2, the spectrum is
increasing compared to the linear spectrum given by the parameter
values related to subject 1.

Using the new scales $\xi_n=Z_n-V_gT_n$ and $\tau_n=T_n$, with
velocity
\begin{equation}\label{eq15}
V_{g}=\frac{D_{0}\sin(q)}{\omega},
\end{equation}
we finally obtain the complex Ginzburg-Landau equation
\begin{equation}\label{eq16}
i\frac{{\partial F_{1}}}{{\partial {\tau _2}}} + P\frac{{{\partial
^2}F_{1}}}{{\partial \xi _1^2}} + (Q_{1}+iQ_{2}){\left| F_{1}
\right|^2}F_{1} + i\gamma F_{1} = 0,
\end{equation}
where
\begin{eqnarray}
\begin{split}
&P=\frac{\omega^{2}D_{0}\cos(q)-D_{0}^{2}\sin^{2}(q)}{4\omega^{3}},\\
&Q_{1}=\frac{1}{\omega}\Big[\frac{2\lambda_{1}D_{2}\sin^{2}(\frac{q}{2})+\lambda_{1}^{2}}{\Omega_{0}^{2}}
+\frac{\omega^{2}\nu_{1}^{2}-3\lambda_{1}^{2}
+8\lambda_{1}^{2}\sin^{2}(\frac{q}{2})}{12\Omega_{0}^{2}+64D_{0}\sin^{4}(\frac{q}{2})}\Big],\\
&Q_{2}=\frac{-\lambda_{1}\nu_{1}}{2\Omega_{0}^{2}}+\frac{3\lambda_{1}\nu_{1}-4D_{2}\nu_{1}\sin^{2}(\frac{q}{2})}{12\Omega_{0}^{2}
+64D_{0}\sin^{4}(\frac{q}{2})},\\
&\gamma  = \frac{\nu_{0}}{4} + D_{1}\sin^{2}(\frac{q}{2}).
\end{split}
\end{eqnarray}

 During the last three decades, the CGL equation and
its modified versions have drawn tremendous attention. These
equations describe a variety of physical phenomena in optical
waveguides and fibers, plasmas, Boose Einstein condensation, phase
transitions, open flow motions,  bimolecular dynamics, spatially
extended non equilibrium systems, etc \cite{Hasegawa}. In the
present research work, the CGL equation is an equation describing
the evolution of modulated hormonal waves in a diffusive coupled
pancreatic islet $\beta$-cells model. This result suggests that, the
insulin propagates within the islet $\beta$-cells using both time
and space domains in order to regulate glucose level. In another
regard, oscillations of insulin secretion which are likely caused by
intrinsic $\beta$-cell mechanisms generate a spatiotemporal dynamics
of insulin between cells as modified by exogenous signals such as
hormonal and neuronal input.

We have represented in Fig. 2. the variations of constants $P$,
$Q_{1}$, $Q_{2}$, $\gamma$ and of the product $PQ_{1}$ with respect
to the wave vector $q$. The parameter values are those in the Table
1. It is observed for both subjects that the coefficients are
positive for small values of the wave vector $q$. The nonlinearity
coefficients have very small values. It is observed also that except
the dissipative coefficient $\gamma$, all the other coefficients
decrease with the increasing of the wave vector. It is well known
that the complex Ginzburg-Landau equation has as the modulational
instability criterion for the plane waves $P_{1}Q_{1}+P_{2}Q_{2}>0$
($P_{1}$ and $P_{2}$ are the real and the imaginary parts of the
dispersion coefficient, respectively) for which the plane waves in
the system are unstable. This relation is known as the Lange and
Newell's criterion \cite{Lange}. However, in this work the imaginary
part of the dispersion coefficient is equal to zero, then the Lange
and Newell's criterion reduces to $P_{1}Q_{1}>0$ known as the
Benjamin-Feir instability criterion \cite{Benjamin}. According to
this instability criterion, for $PQ_{1}>0$ wave planes in the system
are unstable while for $PQ_{1}<0$, they are stable. Since the manner
with which hormonal waves propagate in the system does not depend of
the stability criterion, one can expect to find in the system
spatiotemporal modulated wave solutions for any wave carrier whose
wave vector is in the positive range of $PQ_{1}$.

\subsection{Nonlinear solution of the equation of motion}

The solutions of nonlinear partial differential equations constitute
a crucial factor in the progress of nonlinear dynamics and are a key
access for the understanding of various biological phenomena. Many
analytical investigations have been carried out to find the envelope
soliton solutions of these equations which are localized waves with
particle like behavior i.e., preserving their forms in space or in
time or both in space and time resulting in spatial, temporal or
spatiotemporal solitons, respectively \cite{Shwetanshumala}. We
assume that the form of the envelope soliton solution of Eq. (16)
has the form of the one proposed by Pereira and Stenflo
\cite{Stenflo}, and Nozaki and Bekki \cite{Nozaki}
\begin{equation}\label{18}
F_{1}(\xi_{1},\tau_{2})=
\frac{Ae^{\phi}}{1+e^{(\phi+\phi^{*})^{(1+i\mu)}}}.
\end{equation}

The real part $F_{1r}$ and the imaginary part $F_{1i}$ of
$F_{1}(\xi_{1},\tau_{2})$ are given respectively by
\begin{equation}\label{19}
F_{1r}(\xi_{1},\tau_{2})=A\Big[\frac{e^{-\phi}+\cos(2\mu
\phi)e^{\phi}}{2(\cosh(2\phi)+\cos(2\mu \phi))}\Big] \,\,\,\,\,
\mathrm{and} \,\,\,\,\ F_{1i}(\xi_{1},\tau_{2})=
-A\Big[\frac{\sin(2\mu \phi)e^{\phi}}{2(\cosh(2\phi)+\cos(2\mu
\phi))}\Big],
\end{equation}
where $\phi=q\xi_{1}-\omega\tau_{2}$,
$\mu=-\beta\pm\sqrt{2+\beta^2}$ and $\beta=-\frac{3Q_{1}}{2Q_{2}}$.

Using the solution of $F_{1}$ given by Eq. (19) and from Eq. (4), we
have
\begin{equation}\label{20}
y=2\epsilon(F_{1r}\cos\theta-F_{1i}\sin\theta)
+\epsilon^{2}[F_{0}+2(F_{2r}\cos2\theta-F_{2i}\sin
2\theta)]+O(\epsilon^{3}),
\end{equation}
where $F_{2r}$ and $F_{2i}$ are the real and imaginary parts of
$F_{2}$, respectively such that
\begin{equation}\label{21}
F_{2r}=c_{1}(F_{1r}^{2}-F_{1i}^{2})+2c_{2}F_{1r}F_{1i} \,\,\,\,\,
\mathrm{and} \,\,\,\,\,
F_{2i}=c_{2}(F_{1i}^{2}-F_{1r}^{2})+2c_{1}F_{1r}F_{1i},
\end{equation}
with
\begin{equation}\label{22}
c_{1}=
\frac{\lambda_{1}-4D_{2}\sin^{2}(\frac{q}{2})}{3\Omega_{0}^{2}+16D_{0}\sin^{4}(\frac{q}{2})}
\,\,\,\,\, \mathrm{and} \,\,\,\,\, c_{2}=
\frac{\omega\nu_{1}}{3\Omega_{0}^{2}+16D_{0}\sin^{4}(\frac{q}{2})}.
\end{equation}

Inserting Eqs. (19) and (21) into Eq. (20), we obtain for the
insulin dynamics the following solution
\begin{equation}\label{27}
\begin{split}
y_{n}(t)=&\epsilon
A\Big[\frac{\cos(\theta_{n}-2\alpha\phi_{n})e^{\phi_{n}}+\cos\theta_{n}e^{-\phi_{n}}}{(\cosh2\phi_{n}+\cos2\alpha\phi_{n})}\Big]
+\epsilon A^{2}\Big[-\frac{\lambda_{1}}{\Omega_{0}^{2}(\cosh2\phi_{n}+\cos2\alpha\phi_{n})}\Big]\\
&+\epsilon^{2} A^{2}\Big[(c_{1}\cos2\theta_{n}+c_{2}\sin2\theta_{n})
\times \Big(\frac{2\cos2\alpha\phi_{n} +\cos4\alpha\phi_{n}
e^{2\phi_{n}}+e^{-2\phi_{n}}}{(\cosh2\phi_{n}+\cos2\alpha\phi_{n})}\Big)\Big]\\
&+\epsilon A^{2}\Big[(c_{1}\sin2\theta_{n}-c_{2}\cos2\theta_{n})
\times \Big(\frac{2\sin2\alpha\phi_{n} +\sin4\alpha\phi_{n}
e^{2\phi_{n}}}{(\cosh2\phi_{n}+\cos2\alpha\phi_{n})}\Big)\Big],
\end{split}
\end{equation}
where $\phi_{n}=\epsilon q(n-V_{g}t)-\omega\epsilon^{2}t$.

 In Fig. 3, we have represented the evolution of the
solution at different times according to the parameter values
related to the two first healthy subjects of Ref. \cite{Gaetano}. As
we observe in this figure, the solution is well a localized
modulated solution, and its propagates structurally stable. As
interestingly remarked in the present work, the modulated solution
involving in the system appears in the form of a breather-like
coherent structure and it propagates with the same dynamics for the
different parameter values related to the two healthy subjects. This
assumption can lead to the conclusion that the insulin propagates in
pancreatic islet $\beta$-cells using localized modulated solitonic
waves.

Let us recall that the localized modulated oscillations obtained in
this work are involved in many other biophysical systems. Under
certain conditions, they can move and transport energy along the
system \cite{Mvogo,Peyrard,Zdravkovi}.  As recently demonstrated in
\cite{Mvogo}, breathing modes are also responsible of energy sharing
between $\alpha$-polypeptide coupled chains. Also, localized
oscillations can be precursors of the bubbles that appear in thermal
denaturation of DNA and they have been shown to describe the
breaking of the hydrogen bonds linking two bases \cite{Peyrard}. It
has been also shown that these localized oscillations can move along
microtubule systems \cite{Zdravkovi}. Then, breathers should be
understood as triggering signal for the motor proteins to start
moving as interestingly find also in this work.

In Fig. 4, we have increased the value of $\epsilon$ from
$\epsilon=0.077$ to $\epsilon=0.099$, with the same parameter values
used in Fig. 3. It clearly reveals in this figure the influence of
small perturbation in the dynamics of the hormonal wave. One can
easily see that for both subjects the solutions still remain the
breather excitations. However these breathers are now represented by
modulated solitons so that the envelopes cover less oscillations of
the carrier wave as those observed in Fig. 3. It is also observed
that the amplitude of the wave has increased.

\section{Conclusion}
\label{sec4}

We have studied the intercellular insulin dynamics in an array of
diffusively coupled pancreatic islet $\beta$-cells. The model was
formulated by a system of discrete ordinary differential equations
where the cells are connected via gap junction coupling. Motivated
with often non-observance clinical effects due to some pathological
diseases \cite{Diabete1}, the work was devoted to derive a clear
analytical solution describing the insulin dynamics in pancreatic
islet $\beta$-cells. Applying a powerful perturbation technique, we
have found that the complex Ginzburg-Landau equation is the equation
which describes the insulin dynamics. It has been revealed that the
solution of the hormonal wave is well a localized modulated
solitonic wave called breather. In another regard, the breather has
been revealed as mechanically important in other biophysical systems
such as collagen \cite{Mvogo}, DNA \cite{Peyrard}, microtubule
\cite{Zdravkovi} systems.  The correlation with the present work may
indicate an important role of breathers and other nonlinear
excitations in the dynamics of pancreatic islet $\beta$-cells. In a
forthcoming work, we intend to investigate long-range effect, since
intercellular waves travel also with non contacting cells
\cite{Mvogo2,Kepseu} indicating the long-range interaction in the
system.


\section*{Acknowledgments}

A. Mvogo acknowledges the invitation of the African Institute for
Mathematical Sciences (AIMS). A. Tambue was supported by the Robert
Bosch Stiftung through the AIMS ARETE chair programme.


\newpage

\begin{figure}[h!]
\centering
\includegraphics[width=3in]{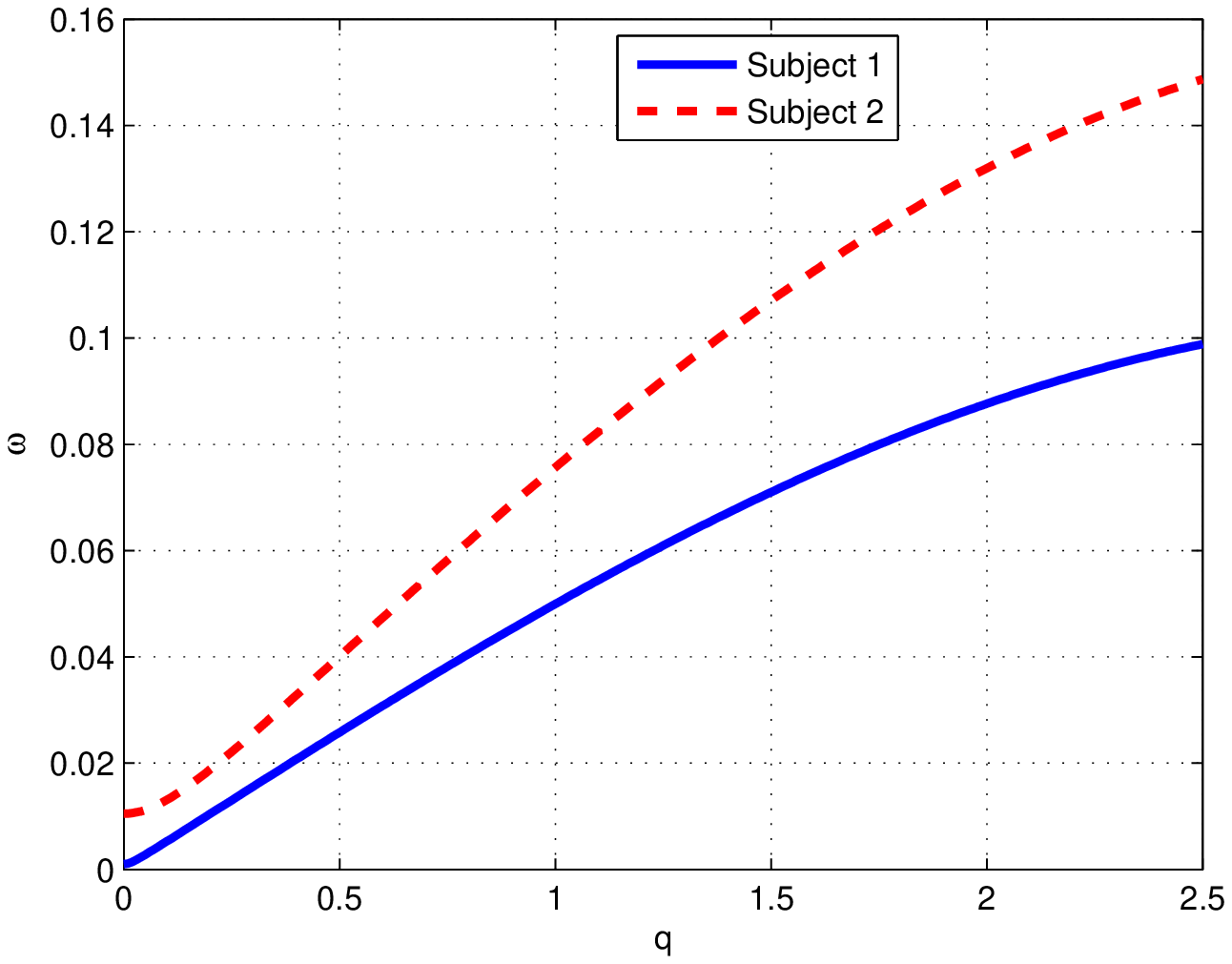}
\caption{(Color online) The dispersion relation of the hormonal
wave. Parameter values are: $D=0.12$ and
$\Omega_{0}^{2}=a_{1}b_{2}$. Subject 1: $a_{1}=0.0226$,
$b_{2}=0.0437$. Subject 2: $a_{1}=0.0509$,
$b_{2}=0.2062$.}\label{fig1}
\end{figure}

\begin{figure}[h!]
\centering
\includegraphics[width=2.5in]{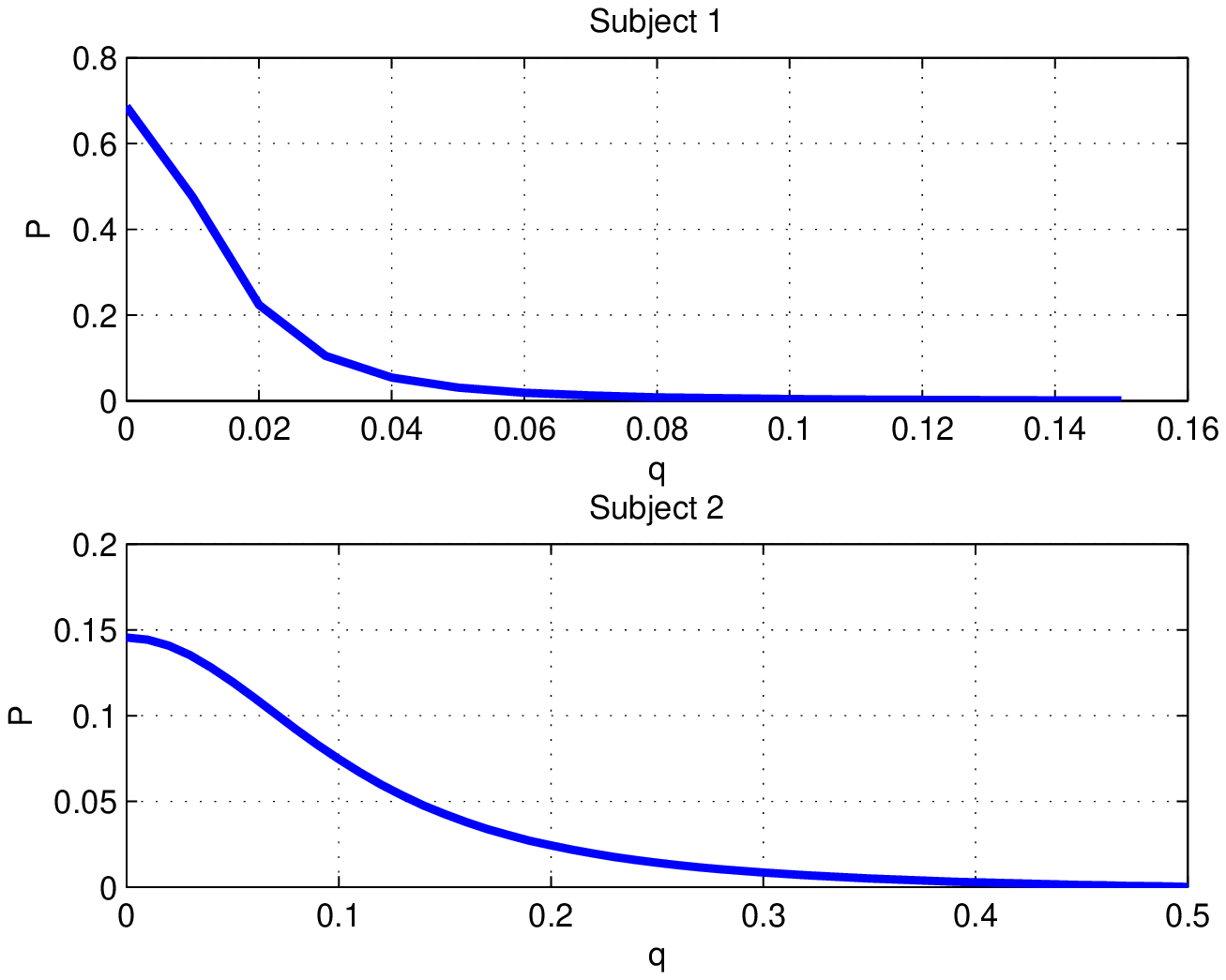}(a)
\includegraphics[width=2.5in]{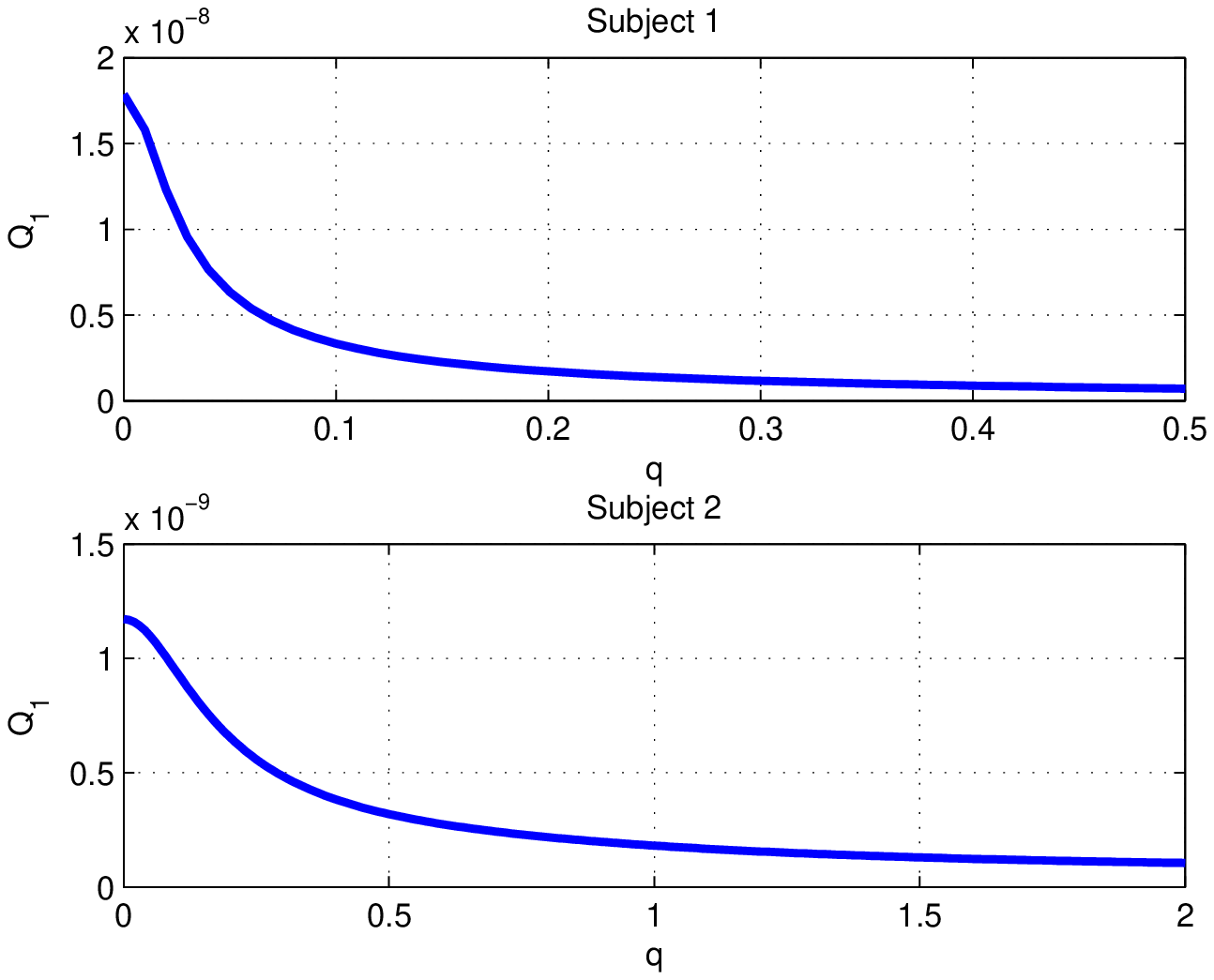}(b)
\includegraphics[width=2.5in]{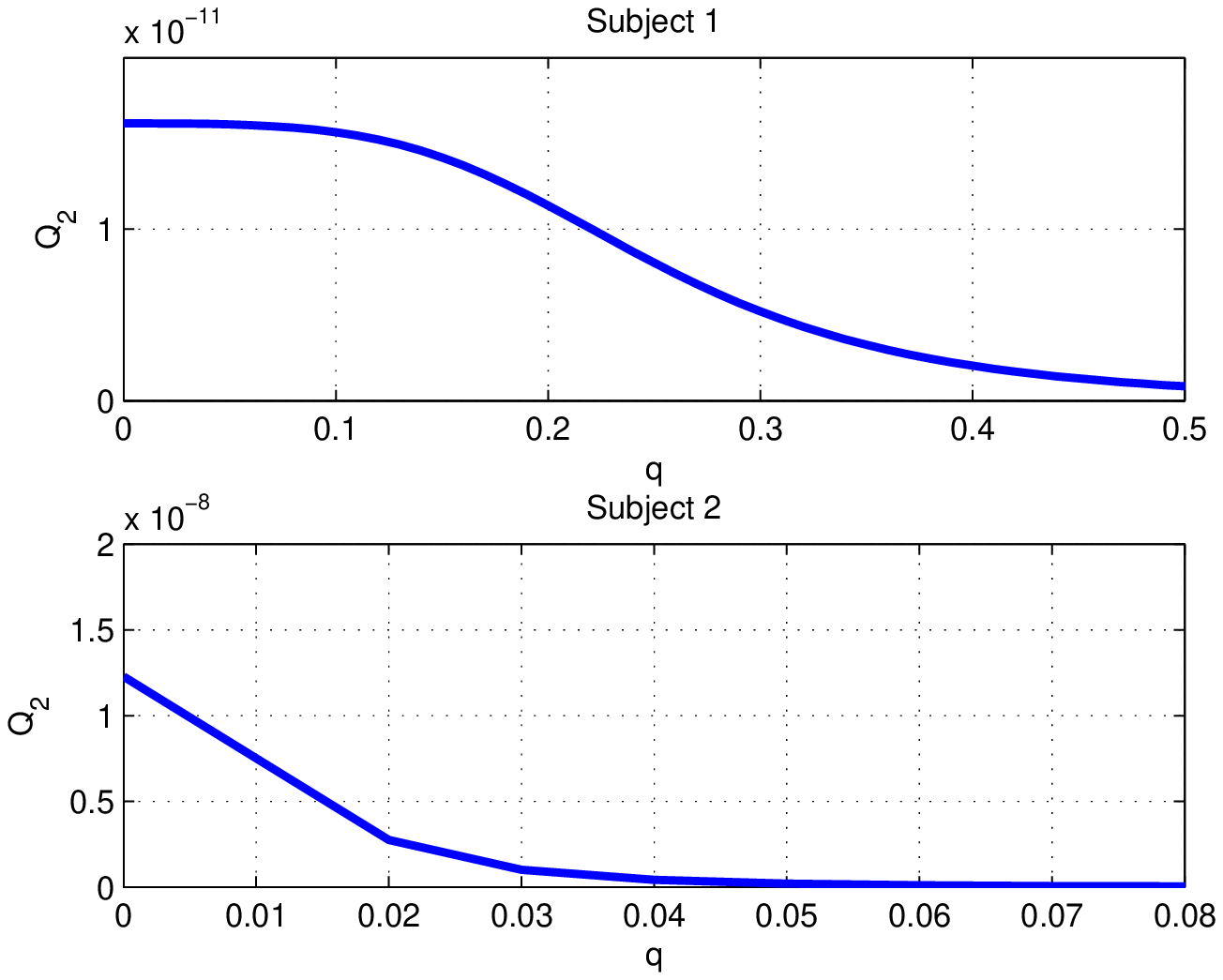}(c)
\includegraphics[width=2.5in]{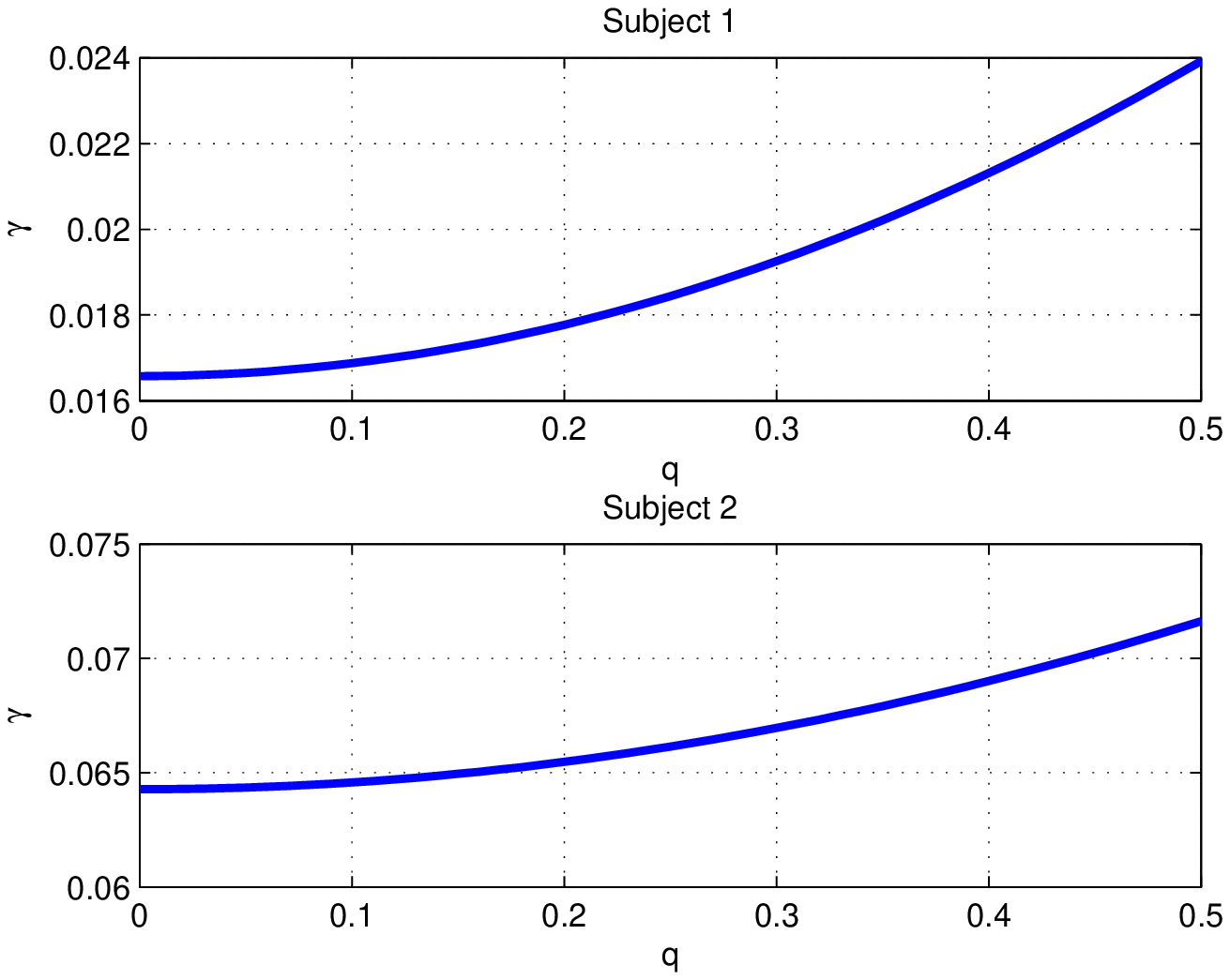}(d)
\includegraphics[width=3.5in]{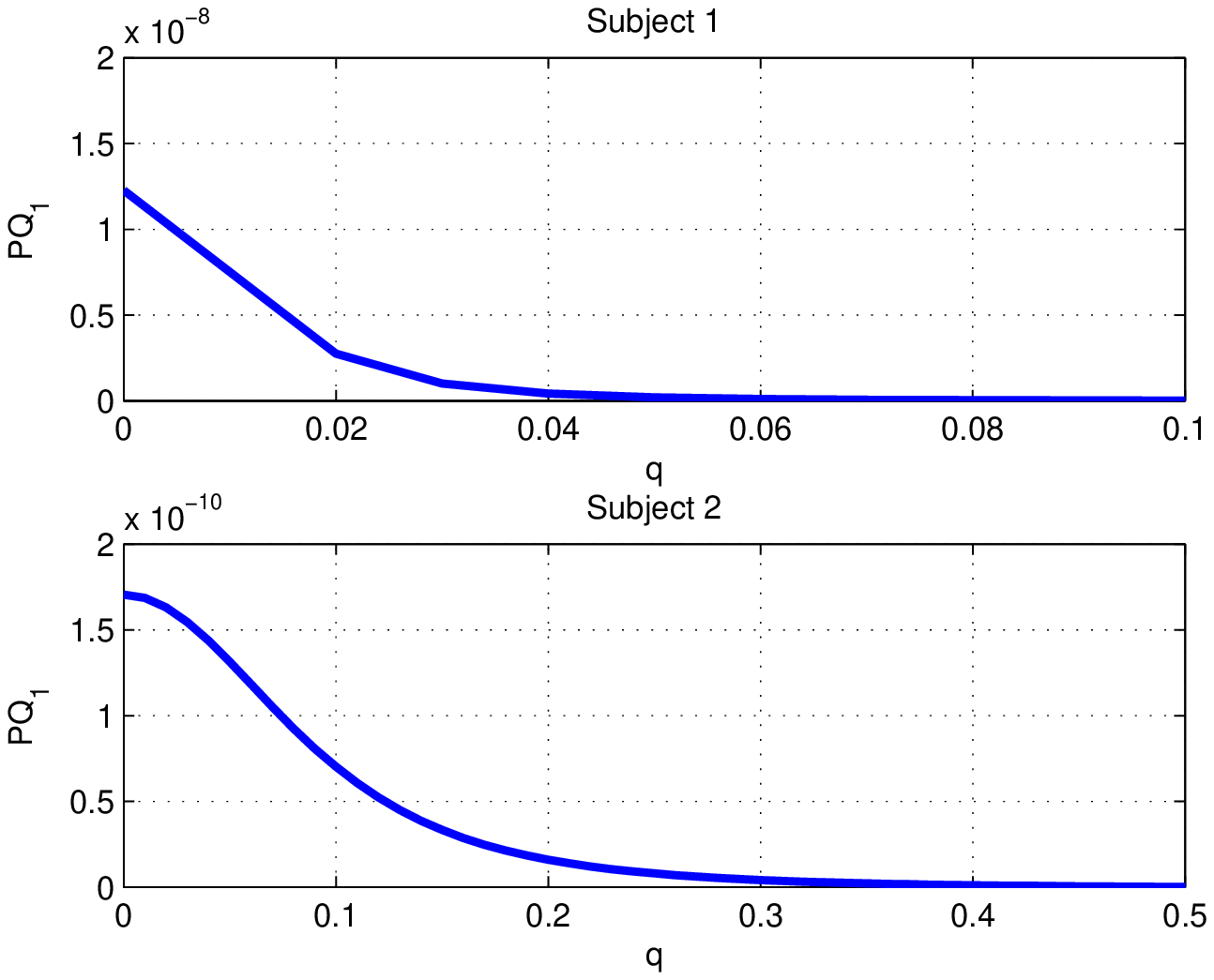}(e)
\caption{(Color online) Variations of coefficients (a) $P$, (b)
$Q_{1}$, (c) $Q_{2}$, (d) the product $PQ_{1}$ as a function of the
wave vector $q$ of the carrier wave. The parameter values are given
in Table 1. Subject 1: $a_{1}=0.0226$, $a_{2}=3.8\times10^{-8}$,
$b_{1}=0.0022$, $b_{2}=0.0437$. Subject 2: $a_{1}=0.0509$,
$a_{2}=1.29\times10^{-7}$, $b_{1}=0.0096$, $b_{2}=0.2062$. With
$D=0.12$.}\label{fig2}
\end{figure}

\begin{figure}[h!]
\centering
\includegraphics[width=3in]{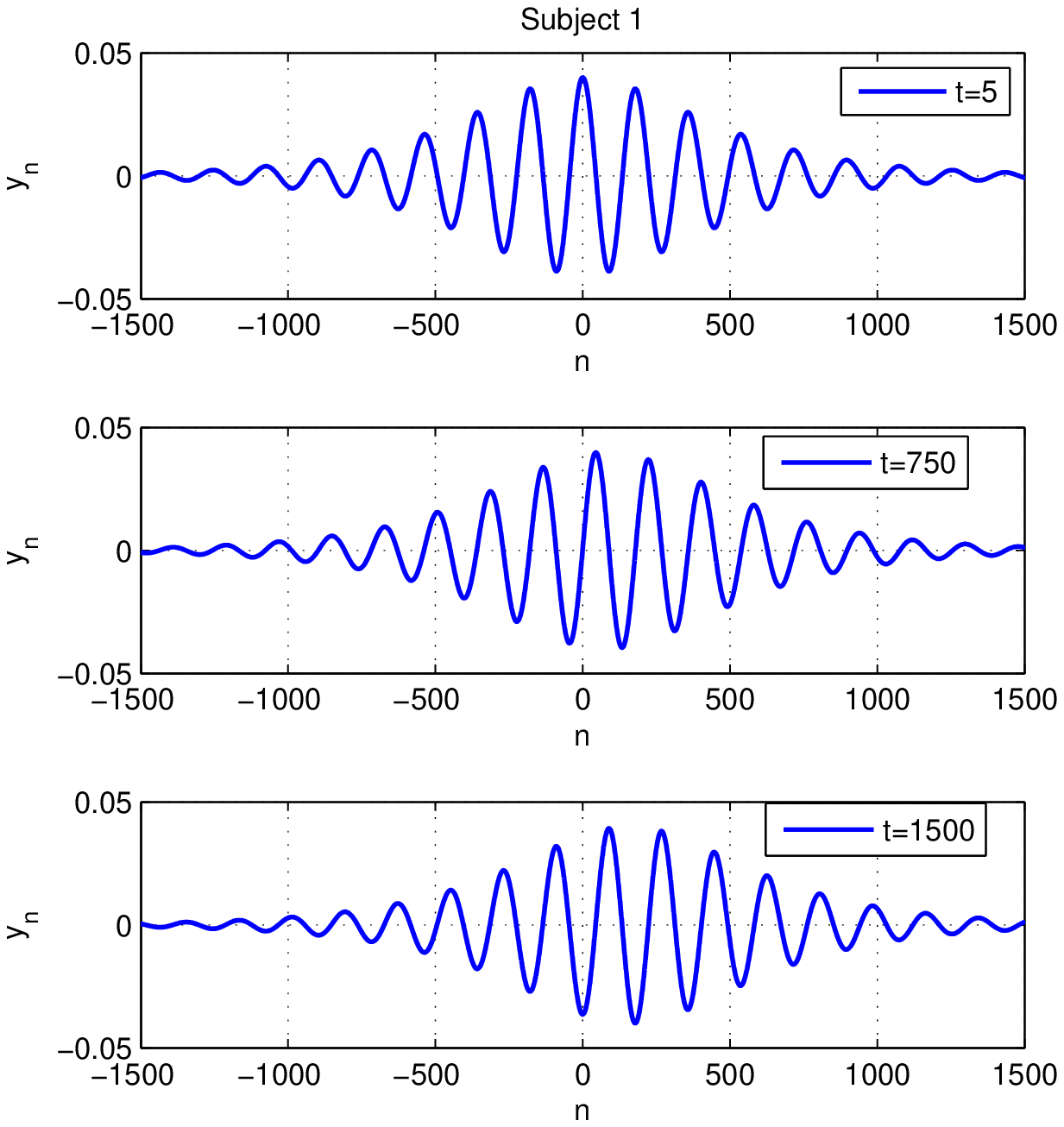}(a)
\includegraphics[width=3in]{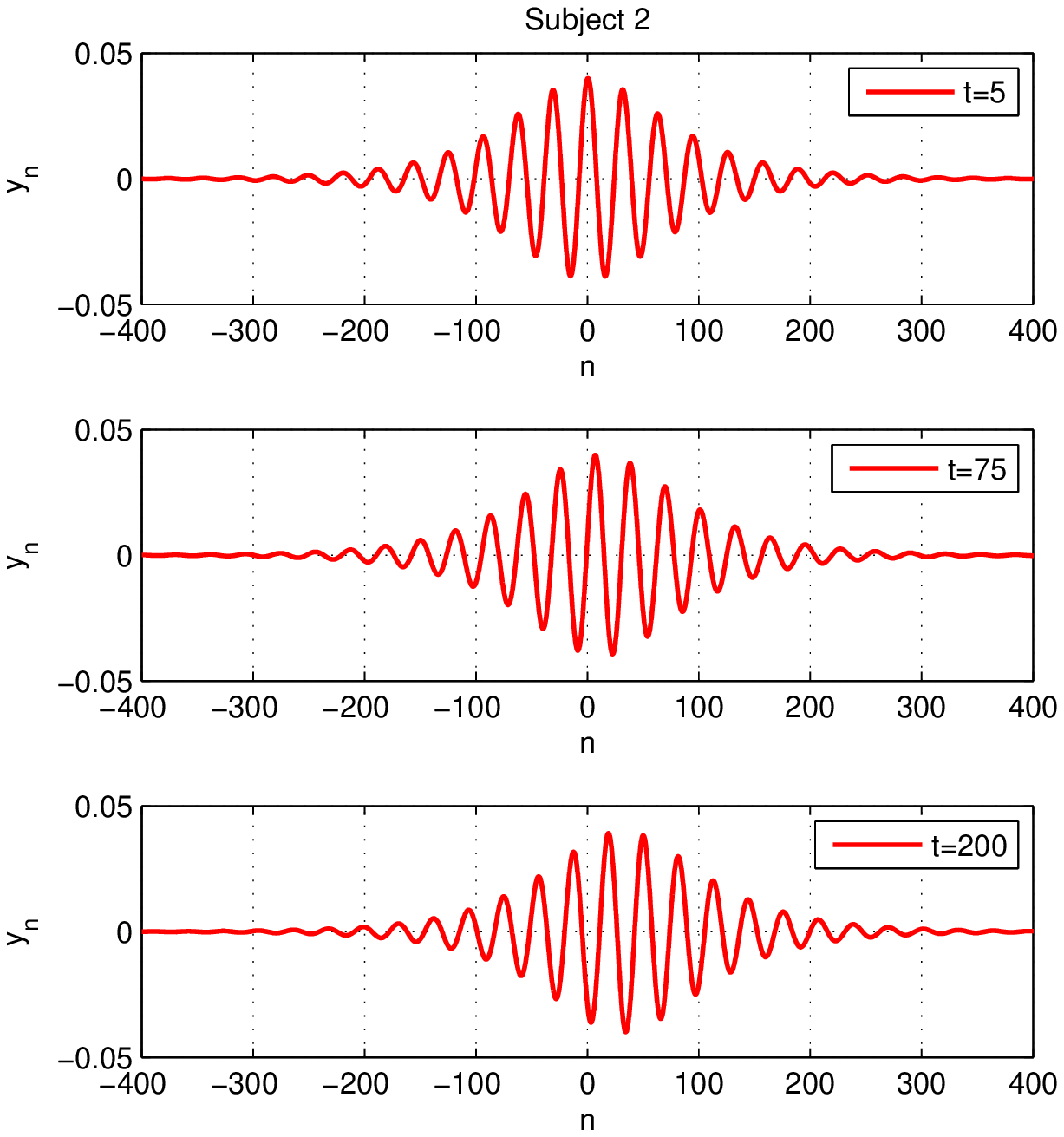}(b)
\caption{(Color online) The solution $y_{n}$  as a function of the
position at different times. The parameter values are: Subject 1:
$D=0.12$, $q=0.035$, $\epsilon=0.077$, $a_{1}=0.0226$,
$b_{1}=0.0022$, $a_{2}=3.8\times10^{-8}$, $b_{2}=0.0437$. Subject 2:
$D=0.12$, $q=0.2$, $\epsilon=0.077$, $a_{1}=0.0509$,
$a_{2}=1.29\times10^{-7}$, $b_{1}=0.0096$,
$b_{2}=0.2062$.}\label{fig3}
\end{figure}

\begin{figure}[h!]
\centering
\includegraphics[width=2.5in]{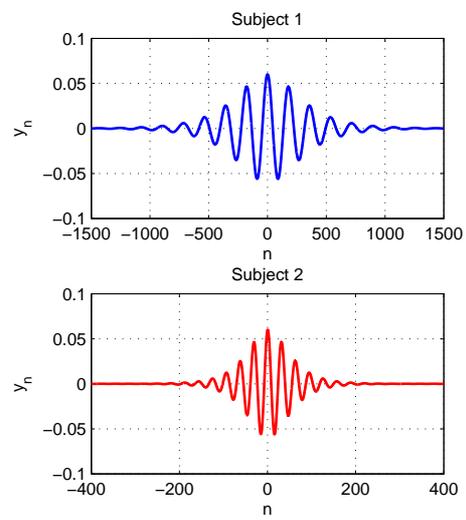}
\caption{(Color online) Effects of small perturbation on the
hormonal modulated wave.  $\epsilon=0.099$ and $t=5$. The parameter
values are the same as in Fig.3.}\label{fig4}
\end{figure}

\end{document}